\documentclass[english,aps,reprint]{revtex4-1}

\usepackage[T1]{fontenc}
\usepackage[latin9]{inputenc}
\usepackage{amsmath}
\usepackage{amssymb}
\usepackage{graphicx}

\makeatletter
\usepackage{dcolumn}
\usepackage{bm}
\usepackage{times}
\usepackage[top=2cm,bottom=2cm,left=1.6cm,right=1.6cm]{geometry}

\makeatother

\usepackage{babel}
\begin{document}
\title{Moment-Fourier approach to parallel closures and transport for a toroidally
confined multi-ion plasma}
\author{Jeong-Young Ji}
\affiliation{Nuclear Research Institute for Future Technology and Policy, Seoul
National University, Seoul 08826, Republic of Korea}
\email{j.y.ji@snu.ac.kr}

\begin{abstract}
We develop a multi-ion closure framework for tokamak plasmas by solving
the drift-kinetic equation using the moment-Fourier method. The model
extends our previous single-ion formulation to plasmas containing
multiple ion species and consistently incorporates inter-species collisional
effects in the moment system. As a representative case, we consider
a deuterium--carbon plasma, assuming a common zeroth-order ion temperature
on each flux surface while allowing the electron temperature to differ.
The first-order fluid moments, namely density, temperature, and parallel
flow velocity, are computed for electrons, deuterium ions, and carbon
ions. The framework enables a direct comparison of the first-order
parallel flow velocities of deuterium and carbon ions, thereby providing
a quantitative basis for interpreting charge-exchange spectroscopy
measurements and for systematic multi-ion neoclassical transport studies.
\end{abstract}
\maketitle

\section{Introduction}

Neoclassical transport theory provides a fundamental framework for
understanding particle, heat, and momentum transport in magnetically
confined plasmas due to Coulomb collisions in an inhomogeneous magnetic
field~\citep{Galeev1968S,Rutherford1970,Hinton1976H,Hirshman1981S,Chang1982H,Helander2002S,Hazeltine2003M}.
The drift kinetic equation serves as the foundation for neoclassical
transport calculations, and numerous numerical methods have been developed
to solve it~\citep{Belli2008C,Belli2009C,Belli2011C,Landreman2012E}.
An analytic approach using the moment-Fourier method has been developed
for ion-only plasmas~\citep{Wong2011,Ji2023HSN} and recently extended
to electron-ion plasmas~\citep{Ji2025LNH}, enabling systematic derivation
of both transport and closure relations with accurate Coulomb collision
effects.

Multi-ion plasmas are ubiquitous in tokamak experiments, where impurity
ions such as carbon play an important role in plasma transport and
rotation. In particular, charge-exchange recombination spectroscopy
(CXRS) provides direct measurements of impurity ion flow velocities~\citep{Chrystal2012},
offering an experimental window into neoclassical transport. However,
a systematic analytic framework for multi-ion neoclassical transport
and closures based on accurate analytic Coulomb collision operators~\citep{Ji2006H,Ji2020SH}
has not been available.

In this work, we extend the moment-Fourier method to multi-ion plasmas.
As a representative case, we consider a deuterium--carbon--electron
(D-C-e) plasma, incorporating inter-species collisional effects consistently
in the moment system. The first-order fluid moments --- density,
temperature, and parallel flow velocity --- are computed for all
three species. The difference in parallel flow velocities between
deuterium and carbon ions is examined as a function of plasma parameters,
providing a quantitative basis for interpreting CXRS measurements
and for systematic multi-ion neoclassical transport studies.

\section{Fourier-moment approach to drift kinetic equation}

In this section, we present a system of parallel moment equations
for a multi-ion plasma that generalizes the single-ion system given
in sections 2 and 3 of Ref.~\citep{Ji2025LNH}. The key distinction
from the single-ion and electron-ion cases lies in the collision operator
$C_{a1}$ {[}Eq.~\eqref{C1}{]}, which now includes inter-species
contributions from all pairs of species in addition to self-species
collisions. These inter-species collision terms couple the moment
equations for different ion species, requiring all species to be treated
simultaneously in a unified linear system.

Although the formalism can be developed for a general magnetic field,
in the present work we consider an axisymmetric circular magnetic
field $\mathbf{B}=I(\psi)\nabla\zeta+\nabla\zeta\times\nabla\psi$
with $B=B_{0}/(1+\epsilon\cos\theta)$, where $\psi$ is a poloidal
flux coordinate, $\zeta$ is the toroidal angle, and $\theta$ is
the poloidal angle. In the formalism, we assume the zeroth-order distribution
function to be Maxwellian, 
\begin{equation}
f_{a0}=\frac{n_{a0}}{\left(2\pi T_{a0}/m_{a}\right)^{3/2}}e^{-w_{a}/T_{a0}}=\frac{n_{a0}}{\pi^{3/2}v^{3}_{a0}}e^{-s^{2}_{a}},\label{fa0}
\end{equation}
where $n_{a0}$ and $T_{a0}$ are flux functions representing the
zeroth-order density and temperature, $v_{a0}=\sqrt{2T_{a0}/m_{a}}$,
and $s_{a}=v/v_{a0}$. Then we solve the following first-order drift
kinetic equation
\begin{equation}
s_{a\|}\frac{\partial}{\partial\theta}f_{a1}-\frac{B}{B^{\theta}v_{a0}}C_{a1}=s_{a\|}\frac{\partial}{\partial\theta}F_{a}+\frac{Bs_{a\|}E_{1\|}}{B^{\theta}T_{a0}}f_{a0},\label{dke1}
\end{equation}
where
\begin{align}
F_{a} & =-\frac{Iv_{\|}}{\Omega_{a}}\left[\frac{\mathrm{\mathrm{d}}\ln p_{a0}}{\mathrm{\mathrm{d}}\psi}+\left(s^{2}_{a}-\frac{5}{2}\right)\frac{\mathrm{\mathrm{d}}\ln T_{a0}}{\mathrm{\mathrm{d}}\psi}+\frac{q_{a}}{T_{a0}}\frac{\mathrm{\mathrm{d}}\phi_{0}}{\mathrm{\mathrm{d}}\psi}\right]f_{a0},\label{Fa:}
\end{align}

and
\begin{equation}
C_{a1}=\sum_{b}\left[C(f_{a1},f_{b0})+C(f_{a0},f_{b1})\right]\label{C1}
\end{equation}
with $C$ denoting the Landau collision operator. Here $\phi_{0}(\psi)$
is the zeroth-order electric potential, and $E_{1\|}$ is the parallel
component of the first-order electric field, which provides an independent
drive for parallel flows in addition to the neoclassical drives contained
in $F_{a}$ {[}Eq.~\eqref{Fa:}{]}.

In the moment-Fourier formulation, the first-order distribution function
$f_{a1}$ is expanded as
\begin{equation}
f_{a1}=f_{a0}\sum^{\mathtt{L}-1}_{l=0}\sum^{\mathtt{K}-1}_{q=0}\sum^{\mathtt{F}-1}_{m=0}M^{lq}_{a(m)}\hat{P}^{lq}_{a}\varphi_{(m)},\label{f:lqm}
\end{equation}
where $\hat{P}^{lq}_{a}=s^{l}_{a}P_{l}(v_{\|}/v)\,L^{(l+1/2)}_{q}(s^{2}_{a})/\sqrt{\bar{\sigma}_{lq}}$,
$s_{a}=v/v_{a0}$, $\bar{\sigma}_{lq}$ is the normalization constant
for a Legendre polynomial $P_{l}$ and an associated Laguerre polynomial
$L^{(l+1/2)}_{q}$; and $\varphi_{(0)}=1$, $\varphi_{(1)}=\varphi_{(1-)}=\sin\theta$,
$\varphi_{(2)}=\varphi_{(1+)}=\cos\theta$, $\varphi_{(3)}=\varphi_{(2-)}=\sin2\theta$,
$\varphi_{(4)}=\varphi_{(2+)}=\cos2\theta$, $\cdots$. The expansion
coefficients $M^{lq}_{a(m)}$ are the $m$-th Fourier component of
the generalized velocity-space moment of $f_{a1}$. The lowest-order
moments are identified with the first-order fluid variables: $M^{00}_{a}=n_{a1}/n_{a0}$,
$M^{01}_{a}=-\sqrt{3/2}\,T_{a1}/T_{a0}$, $M^{10}_{a}=\sqrt{2}\,u_{a}/v_{a0}$,
$M^{11}_{a}=-\sqrt{4/5}\,h_{a\|}/v_{a0}p_{a0}$, and $M^{20}_{a}=\sqrt{3/4}\,\pi_{a\|}/p_{a0}$,
where $n_{a1}$, $T_{a1}$, $u_{a}$, $h_{a\|}$, and $\pi_{a\|}$
are the first-order density, temperature, parallel flow velocity,
parallel heat flux density, and parallel viscosity, respectively.

Taking moments of Eq.~\eqref{dke1} with $\hat{P}^{lp}_{a}$ yields
a set of coupled ordinary differential equations (ODEs) in the poloidal
angle $\theta$. All $\theta$-dependent quantities --- the moments
and the magnetic-field functions --- are further expanded in Fourier
series in $\theta$, converting the ODE system to a block-structured
set of linear algebraic equations for the Fourier coefficients $\{M^{lq}_{a(m)}\}$.

For a multi-ion plasma, the first-order collision term $C_{a1}$ {[}Eq.~\eqref{C1}{]}
couples the moment equations of species $a$ to those of every other
species $b$ through inter-species Coulomb collisions. The resulting
system for species $a$ takes the form
\begin{equation}
\mathsf{D}_{a}\mathbb{M}_{a}-\sum\left(\mathsf{A}_{ab}\mathbb{M}_{a}+\mathsf{B}_{ab}\mathbb{M}_{b}\right)=\mathbb{G}_{a},\label{M:G}
\end{equation}
where $\mathsf{D}_{a}$, $\mathsf{A}_{ab}$, and $\mathsf{B}_{ab}$
are the $\mathtt{LKF}$ dimensional moment-Fourier matrices of $s_{a\|}\partial/\partial_{\theta}$,
$(B/B^{\theta}v_{a0})C(f_{a1},f_{b0})$, and $(B/B^{\theta}v_{a0})C(f_{a0},f_{b1})$,
respectively; and $\mathbb{M}_{a}$ is the column vector of $\{M^{lq}_{a(m)}\}$;
and $\mathbb{G}_{a}$ is the moment-Fourier vector of the right-hand
side of Eq.~\eqref{dke1} ($F_{a}$ and $E_{1\|}$ terms). The full
mass-ratio dependence is retained in all collision matrices without
the small-mass ratio approximation. Although the approximation provides
significantly reduced electron-ion and ion-electron collision matrices,
it cannot be applied to ion-ion collisions consistently in a multi-ion
plasma. Therefore it is convenient to adopt accurate collision operators
for electrons and all ion species uniformly. Systems of the form of
Eq.~\eqref{M:G} for all species are assembled into a single linear
system and solved simultaneously. The resulting first-order density,
temperature, and parallel flow velocity for each species are expressed
as linear combinations of the radial pressure and temperature gradients,
$\mathrm{d}p_{a0}/\mathrm{d}\psi$ and $\mathrm{d}T_{a0}/\mathrm{d}\psi$,
and the first-order parallel electric field $E_{1\|}$. 

The closure relations for the density, temperature, and flow velocity
equations can be obtained by solving Eq.~\eqref{M:G} with the rows
for $\{M^{00}_{a},M^{01}_{a},M^{10}_{a}\}$ moments removed. Due to
the removal of the rows, the corresponding columns appear as closure
drives in addition to $F_{a}$ drive terms. The additional closure
terms are parallel gradients of $T_{a1}$ and $u_{a}$. Since the
$E_{1\|}$ term has appeared only in the $M^{10}_{a}$ equation, it
does not appear in the system of equations for closures. Solving the
system of equations yields the necessary moments to obtain closures,
$\{Q_{a},h_{a\|},R_{a\|},\pi_{a\|}\}$, where $Q_{a}$ and $R_{a\|}$
are collisional heating and friction, respectively. The closure relations
are expressed in terms of fluid variables in Fourier representation.
Once the closure relations have been obtained, they can be combined
with the linearized fluid equations to obtain$n_{a1}$, $T_{a1}$,
and $u_{a}$. It is obvious and we have verified that the fluid quantities
obtained from the full transport system {[}Eq.~\eqref{M:G}{]} agree
with those obtained by combining the closure relations with the linearized
fluid equations. 

Letting $\mathbb{F}_{a}$ denote the Fourier column vector of $n_{a1}/n_{a0}$,
$T_{a1}/T_{a0}$, and $u_{a}/v_{a0}$, the solution can be written
as
\begin{equation}
\mathbb{F}_{a}=\sum_{b}\mathbb{F}^{p}_{ab}\hat{p}_{b0,\psi}+\sum_{b}\mathbb{F}^{T}_{ab}\hat{T}_{b0,\psi}+\sum_{n}\mathbb{F}^{E(m)}_{a}\hat{E}_{(m)},\label{Fa:pTE}
\end{equation}
where
\begin{align*}
\hat{p}_{a0,\psi} & =\frac{I}{q_{a}v_{a0}B_{0}}\left(\frac{1}{n_{a0}}\frac{\mathrm{d}p_{a0}}{\mathrm{d}\psi}+q_{a}\frac{\mathrm{d}\phi_{0}}{\mathrm{d}\psi}\right),\\
\hat{T}_{a0,\psi} & =\frac{I}{q_{a}v_{a0}B_{0}}\frac{\mathrm{d}T_{a0}}{\mathrm{d}\psi},
\end{align*}
and $\hat{E}_{(m)}$ is the $m$-th Fourier component of $\hat{E}=eE_{1\|}B/B^{\theta}T_{\mathrm{e}0}$.
The coefficient column vectors $\mathbb{F}^{p}_{ab}$, $\mathbb{F}^{T}_{ab}$,
and $\mathbb{F}^{E(m)}_{a}$ encode the full neoclassical transport
response of species $a$ to the drives of species $b$, and the electric
field. Then the first-order fluid quantities can be expressed in Fourier
series $\sum_{n}\mathbb{F}_{a(n)}\varphi_{(n)}$ for each drive term
$\hat{p}_{b0,\psi}$, $\hat{T}_{b0,\psi}$, and $\hat{E}_{(m)}$,
where $\mathbb{F}_{a(n)}$ is the $n$-th Fourier coefficient.

\begin{figure*}
\includegraphics[width=0.9\textwidth]{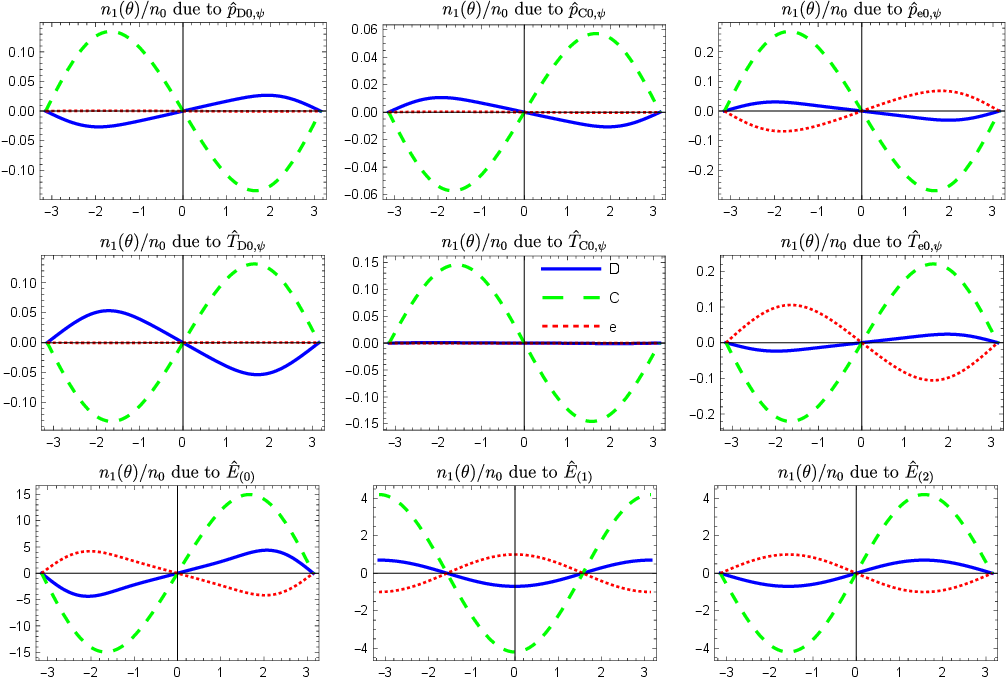}

\caption{First-order density perturbation $n_{1}/n_{0}$ of D (solid, blue),
C (dashed, green), and electrons (dotted, red) driven by the pressure
gradient $\hat{p}_{b0,\psi}$, the temperature gradient $\hat{T}_{b0,\psi}$,
$(b=\mathrm{D},\mathrm{C},\mathrm{e})$, and the electric field $\hat{E}_{(m)}\;(m=0,1,2)$
as functions of the poloidal angle $\theta$. \label{fig:density}}
\end{figure*}
\begin{figure*}
\includegraphics[width=0.9\textwidth]{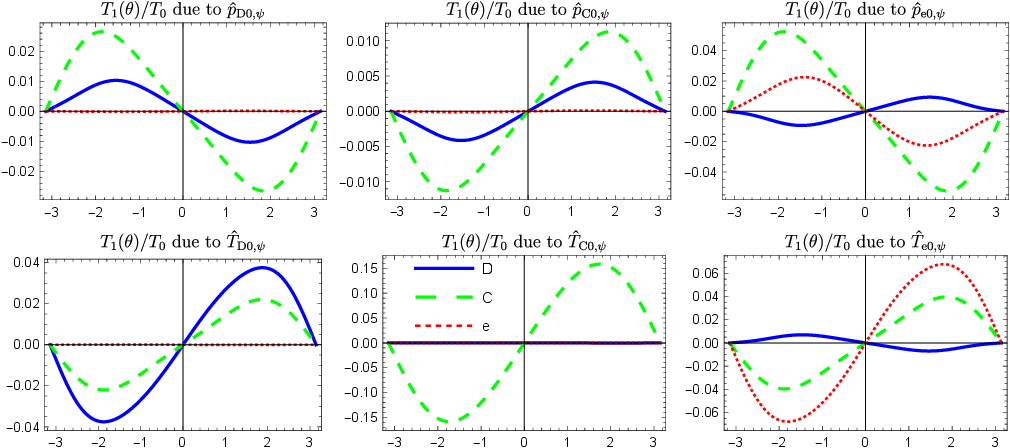}

\caption{First-order temperature perturbation $T_{1}/T_{0}$ of D (solid, blue),
C (dashed, green), and electrons (dotted, red) driven by the pressure
gradient $\hat{p}_{b0,\psi}$ and the temperature gradient $\hat{T}_{b0,\psi}$.
\label{fig:temp}}
\end{figure*}

\begin{figure*}
\includegraphics[width=0.9\textwidth]{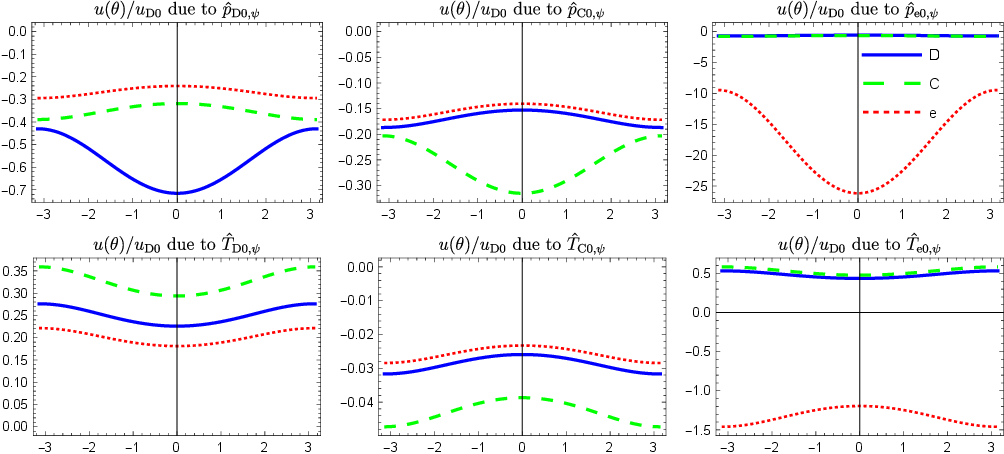}

\caption{Parallel flow velocity $u/u_{\mathrm{D}0}$ of D (solid, blue), C
(dashed, green), and electrons (dotted, red) driven by the pressure
gradient $\hat{p}_{b0,\psi}$ and the temperature gradient $\hat{T}_{b0,\psi}$.
\label{fig:flows}}
\end{figure*}

\section{Results}

\begin{figure}
\includegraphics[width=0.9\columnwidth]{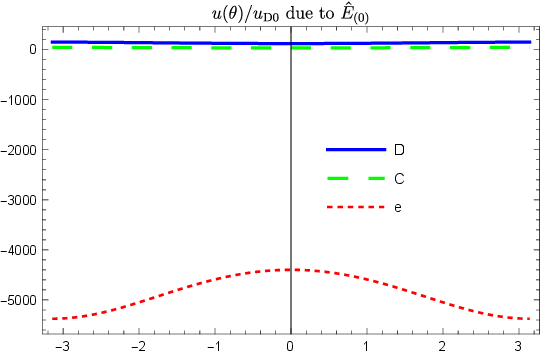}

\caption{Parallel flow velocity $u/u_{\mathrm{D}0}$ of D (solid, blue), C
(dashed, green), and electrons (dotted, red), driven by $E_{1\|}$.\label{fig:flow2E}}
\end{figure}

We apply the framework to a D-C-e plasma with inverse aspect ratio
$\epsilon=0.1$. The electron and common ion temperatures are $T_{\mathrm{e}}=3\,\mathrm{keV}$
and $T_{\mathrm{D}}=T_{\mathrm{C}}=1.4\,T_{\mathrm{e}}$. Densities
are $n_{\mathrm{e}}=2\times10^{19}\,\mathrm{m}^{-3}$, $n_{\mathrm{D}}=0.64\,n_{\mathrm{e}}$,
and $n_{\mathrm{C}}=0.06\,n_{\mathrm{e}}$, which satisfy quasi-neutrality
$n_{\mathrm{D}}+Z_{\mathrm{C}}n_{\mathrm{C}}=n_{\mathrm{e}}$ with
$Z_{\mathrm{C}}=6$. The Knudsen number for the deuterium ions is
$10^{3}$. The linear system \eqref{M:G} is constructed and solved
for$\mathtt{L}=40$,$\mathtt{K}=80$, and$\mathtt{F}=9$. The resulting
profiles of $n_{a1}$, $T_{a1}$, and $u_{a}$ shown in Figs.~\ref{fig:density}--\ref{fig:flow2E}
are reconstructed from the Fourier series $\sum_{n}\mathbb{F}_{a(n)}\varphi_{(n)}$,
retaining $\mathtt{F}=9$ harmonics.

Figures~\ref{fig:density} and~\ref{fig:temp} show the first-order
density and temperature perturbations for D, C, and electrons driven
separately by each neoclassical drive. The first-order density and
temperature perturbations depend on the radial gradients of the zeroth-order
densities and temperatures of all species and on the Fourier components
of the parallel electric field. The contribution of D and C ion drives
to the first-order electron density and temperature is negligible
compared to the electron contribution. Figure~\ref{fig:flows} shows
the parallel flow velocities driven by the pressure and temperature
gradients. The D and C ion flows respond differently to each neoclassical
drive, and the resulting flow velocity difference suggests that competing
neoclassical drives may provide an explanation for the intrinsic rotation
reversal. Figure~\ref{fig:flow2E} shows the parallel flow velocities
driven by the parallel electric field $E_{1\|}$. For a circular magnetic
field, the parallel flow velocity is driven only by the zeroth Fourier
component $\hat{E}_{(0)}$, not by $\hat{E}_{(m)}$ for $m>0$. The
D and C velocities differ noticeably from each other due to the mass
dependence in the inter-species collision matrices. Since C ion flows
are directly measurable by CXRS, these results provide a basis for
inferring D ion flows and for comparisons with experimental data.

\section{Summary}

We have extended the moment-Fourier method~\citep{Ji2023HSN,Ji2025LNH}
to multi-ion plasmas and applied it to a D-C-e plasma. The framework
incorporates inter-species Coulomb collision effects consistently
in the moment-Fourier system without the small mass-ratio approximation,
treating electrons and all ion species on the same footing in a single
linear system. As an example study, we apply the method to a D-C-e
plasma and compute the first-order density, temperature, and parallel
flow velocity for all three species, driven separately by the radial
pressure and temperature gradients and the parallel electric field. 

The D and C ion flows respond differently to each neoclassical drive,
suggesting that competing drives may underlie the intrinsic rotation
reversal. Understanding the flow difference between D and C ions also
provides a basis for inferring D ion flows from CXRS measurements
of C ions. Future work will address bootstrap current studies, intrinsic
rotation analyses, and comparisons with experimental CXRS data.
\begin{acknowledgments}
The research was supported by the Brain Pool program funded by the
Ministry of Science and ICT through the NRF of Korea (RS-2024-00409580).
\end{acknowledgments}


\begin{thebibliography}{17}
\expandafter\ifx\csname natexlab\endcsname\relax\def\natexlab#1{#1}\fi
\expandafter\ifx\csname bibnamefont\endcsname\relax
  \def\bibnamefont#1{#1}\fi
\expandafter\ifx\csname bibfnamefont\endcsname\relax
  \def\bibfnamefont#1{#1}\fi
\expandafter\ifx\csname citenamefont\endcsname\relax
  \def\citenamefont#1{#1}\fi
\expandafter\ifx\csname url\endcsname\relax
  \def\url#1{\texttt{#1}}\fi
\expandafter\ifx\csname urlprefix\endcsname\relax\def\urlprefix{URL }\fi
\providecommand{\bibinfo}[2]{#2}
\providecommand{\eprint}[2][]{\url{#2}}

\bibitem[{\citenamefont{Galeev and Sagdeev}(1968)}]{Galeev1968S}
\bibinfo{author}{\bibfnamefont{A.~A.} \bibnamefont{Galeev}} \bibnamefont{and}
  \bibinfo{author}{\bibfnamefont{R.~Z.} \bibnamefont{Sagdeev}},
  \bibinfo{journal}{Sov. Phys. JETP} \textbf{\bibinfo{volume}{26}},
  \bibinfo{pages}{233} (\bibinfo{year}{1968}).

\bibitem[{\citenamefont{Rutherford}(1970)}]{Rutherford1970}
\bibinfo{author}{\bibfnamefont{P.~H.} \bibnamefont{Rutherford}},
  \bibinfo{journal}{Phys. Fluids} \textbf{\bibinfo{volume}{13}},
  \bibinfo{pages}{482} (\bibinfo{year}{1970}), ISSN \bibinfo{issn}{0031-9171}.

\bibitem[{\citenamefont{Hinton and Hazeltine}(1976)}]{Hinton1976H}
\bibinfo{author}{\bibfnamefont{F.~L.} \bibnamefont{Hinton}} \bibnamefont{and}
  \bibinfo{author}{\bibfnamefont{R.~D.} \bibnamefont{Hazeltine}},
  \bibinfo{journal}{Rev. Mod. Phys.} \textbf{\bibinfo{volume}{48}},
  \bibinfo{pages}{239} (\bibinfo{year}{1976}).

\bibitem[{\citenamefont{Hirshman and Sigmar}(1981)}]{Hirshman1981S}
\bibinfo{author}{\bibfnamefont{S.}~\bibnamefont{Hirshman}} \bibnamefont{and}
  \bibinfo{author}{\bibfnamefont{D.}~\bibnamefont{Sigmar}},
  \bibinfo{journal}{Nucl. Fusion} \textbf{\bibinfo{volume}{21}},
  \bibinfo{pages}{1079} (\bibinfo{year}{1981}).

\bibitem[{\citenamefont{Chang and Hinton}(1982)}]{Chang1982H}
\bibinfo{author}{\bibfnamefont{C.~S.} \bibnamefont{Chang}} \bibnamefont{and}
  \bibinfo{author}{\bibfnamefont{F.~L.} \bibnamefont{Hinton}},
  \bibinfo{journal}{Phys. Fluids} \textbf{\bibinfo{volume}{25}},
  \bibinfo{pages}{1493} (\bibinfo{year}{1982}).

\bibitem[{\citenamefont{Helander and Sigmar}(2002)}]{Helander2002S}
\bibinfo{author}{\bibfnamefont{P.}~\bibnamefont{Helander}} \bibnamefont{and}
  \bibinfo{author}{\bibfnamefont{D.~J.} \bibnamefont{Sigmar}},
  \emph{\bibinfo{title}{Collisional Transport in Magnetized Plasmas}}
  (\bibinfo{publisher}{Cambridge University Press, Cambridge},
  \bibinfo{year}{2002}).

\bibitem[{\citenamefont{Hazeltine and Meiss}(2003)}]{Hazeltine2003M}
\bibinfo{author}{\bibfnamefont{R.~D.} \bibnamefont{Hazeltine}}
  \bibnamefont{and} \bibinfo{author}{\bibfnamefont{J.~D.} \bibnamefont{Meiss}},
  \emph{\bibinfo{title}{Plasma Confinement}} (\bibinfo{publisher}{Dover Pub.,
  Inc., New York}, \bibinfo{year}{2003}).

\bibitem[{\citenamefont{Belli and Candy}(2008)}]{Belli2008C}
\bibinfo{author}{\bibfnamefont{E.~A.} \bibnamefont{Belli}} \bibnamefont{and}
  \bibinfo{author}{\bibfnamefont{J.}~\bibnamefont{Candy}},
  \bibinfo{journal}{Plasma Phys. Control. Fusion}
  \textbf{\bibinfo{volume}{50}}, \bibinfo{pages}{095010}
  (\bibinfo{year}{2008}).

\bibitem[{\citenamefont{Belli and Candy}(2009)}]{Belli2009C}
\bibinfo{author}{\bibfnamefont{E.~A.} \bibnamefont{Belli}} \bibnamefont{and}
  \bibinfo{author}{\bibfnamefont{J.}~\bibnamefont{Candy}},
  \bibinfo{journal}{Plasma Phys. Control. Fusion}
  \textbf{\bibinfo{volume}{51}}, \bibinfo{pages}{075018}
  (\bibinfo{year}{2009}).

\bibitem[{\citenamefont{Belli and Candy}(2011)}]{Belli2011C}
\bibinfo{author}{\bibfnamefont{E.~A.} \bibnamefont{Belli}} \bibnamefont{and}
  \bibinfo{author}{\bibfnamefont{J.}~\bibnamefont{Candy}},
  \bibinfo{journal}{Plasma Phys. Control. Fusion}
  \textbf{\bibinfo{volume}{54}}, \bibinfo{pages}{015015}
  (\bibinfo{year}{2011}).

\bibitem[{\citenamefont{Landreman and Ernst}(2012)}]{Landreman2012E}
\bibinfo{author}{\bibfnamefont{M.}~\bibnamefont{Landreman}} \bibnamefont{and}
  \bibinfo{author}{\bibfnamefont{D.~R.} \bibnamefont{Ernst}},
  \bibinfo{journal}{Plasma Phys. Control. Fusion}
  \textbf{\bibinfo{volume}{54}}, \bibinfo{pages}{115006}
  (\bibinfo{year}{2012}).

\bibitem[{\citenamefont{Wong and Chan}(2011)}]{Wong2011}
\bibinfo{author}{\bibfnamefont{S.~K.} \bibnamefont{Wong}} \bibnamefont{and}
  \bibinfo{author}{\bibfnamefont{V.~S.} \bibnamefont{Chan}},
  \bibinfo{journal}{Plasma Phys. Control. Fusion}
  \textbf{\bibinfo{volume}{53}}, \bibinfo{pages}{095005}
  (\bibinfo{year}{2011}).

\bibitem[{\citenamefont{Ji et~al.}(2023)\citenamefont{Ji, Held, Spencer, and
  Na}}]{Ji2023HSN}
\bibinfo{author}{\bibfnamefont{J.-Y.} \bibnamefont{Ji}},
  \bibinfo{author}{\bibfnamefont{E.~D.} \bibnamefont{Held}},
  \bibinfo{author}{\bibfnamefont{J.~A.} \bibnamefont{Spencer}},
  \bibnamefont{and} \bibinfo{author}{\bibfnamefont{Y.-S.} \bibnamefont{Na}},
  \bibinfo{journal}{Plasma Phys. Control. Fusion}
  \textbf{\bibinfo{volume}{65}}, \bibinfo{pages}{035018}
  (\bibinfo{year}{2023}).

\bibitem[{\citenamefont{Ji et~al.}(2025)\citenamefont{Ji, Lee, Na, and
  Held}}]{Ji2025LNH}
\bibinfo{author}{\bibfnamefont{J.-Y.} \bibnamefont{Ji}},
  \bibinfo{author}{\bibfnamefont{M.~U.} \bibnamefont{Lee}},
  \bibinfo{author}{\bibfnamefont{Y.-S.} \bibnamefont{Na}}, \bibnamefont{and}
  \bibinfo{author}{\bibfnamefont{E.~D.} \bibnamefont{Held}},
  \bibinfo{journal}{Plasma Phys. Control. Fusion}
  \textbf{\bibinfo{volume}{67}}, \bibinfo{pages}{045034}
  (\bibinfo{year}{2025}).

\bibitem[{\citenamefont{Chrystal et~al.}(2012)\citenamefont{Chrystal, Burrell,
  Grierson, Groebner, and Kaplan}}]{Chrystal2012}
\bibinfo{author}{\bibfnamefont{C.}~\bibnamefont{Chrystal}},
  \bibinfo{author}{\bibfnamefont{K.~H.} \bibnamefont{Burrell}},
  \bibinfo{author}{\bibfnamefont{B.~A.} \bibnamefont{Grierson}},
  \bibinfo{author}{\bibfnamefont{R.~J.} \bibnamefont{Groebner}},
  \bibnamefont{and} \bibinfo{author}{\bibfnamefont{D.~H.}
  \bibnamefont{Kaplan}}, \bibinfo{journal}{Rev. Sci. Instrum.}
  \textbf{\bibinfo{volume}{83}}, \bibinfo{pages}{10D501}
  (\bibinfo{year}{2012}).

\bibitem[{\citenamefont{Ji and Held}(2006)}]{Ji2006H}
\bibinfo{author}{\bibfnamefont{J.-Y.} \bibnamefont{Ji}} \bibnamefont{and}
  \bibinfo{author}{\bibfnamefont{E.~D.} \bibnamefont{Held}},
  \bibinfo{journal}{Phys. Plasmas} \textbf{\bibinfo{volume}{13}},
  \bibinfo{pages}{102103} (\bibinfo{year}{2006}).

\bibitem[{\citenamefont{Ji et~al.}(2020)\citenamefont{Ji, Spencer, and
  Held}}]{Ji2020SH}
\bibinfo{author}{\bibfnamefont{J.-Y.} \bibnamefont{Ji}},
  \bibinfo{author}{\bibfnamefont{J.~A.} \bibnamefont{Spencer}},
  \bibnamefont{and} \bibinfo{author}{\bibfnamefont{E.~D.} \bibnamefont{Held}},
  \bibinfo{journal}{Plasma Res. Express} \textbf{\bibinfo{volume}{2}},
  \bibinfo{pages}{015013} (\bibinfo{year}{2020}).

\end{thebibliography}
\end{document}